\documentclass[12pt,a4paper]{article}
\usepackage[T1]{fontenc}
\usepackage[backend=bibtex,
            style=numeric-comp,
            maxcitenames=300,
            maxbibnames=500,
            block=none,
            sorting=none
            ]{biblatex}
\usepackage{slashed}
\usepackage{epsfig}
\usepackage{lscape} 
\usepackage{enumerate}
\usepackage{supertabular}
\usepackage{amsmath}
\usepackage{amssymb}
\usepackage{graphicx}
\usepackage{rotating}
\usepackage[toc,page]{appendix}
\usepackage{alphalph}
\usepackage{perpage}
\MakePerPage{footnote}
\usepackage[section] {placeins} 
\DeclareMathAlphabet{\scr}{U}{rsfs}{m}{n}
\newcommand{\bmp}{\begin{minipage}[t]{9cm}}
\newcommand{\bmpz}{\begin{minipage}[t]{5cm}}
\newcommand{\emp}{\end{minipage}}
\numberwithin{equation}{section}
\numberwithin{figure}{section}
\numberwithin{table}{section}
\usepackage{fancybox}
\usepackage{tikz}
\usetikzlibrary{positioning,arrows}
\usetikzlibrary{trees}
\usetikzlibrary{decorations.pathmorphing}
\usetikzlibrary{decorations.markings}
  \definecolor{jblue}  {RGB}{20,50,100}
  \definecolor{npurple}  {RGB} {153, 51, 204}
  \definecolor{wred}   {RGB}{217,0,56}
  \definecolor{white}   {RGB}{255,255,255}  
  \definecolor{korange}   {RGB}{235, 80,  43}
  \definecolor{korange2}   {RGB}{245, 100,  63}
  \definecolor{kyelloworange}   {RGB}{255, 210,  110}
  \definecolor{kyelloworange2}   {RGB}{240, 170,  90}
  \definecolor{kred}   {RGB}{204,  102, 153}
  \definecolor{kpurple}   {RGB}{153,  61, 190}
  \definecolor{kpurplelight}   {RGB}{213,  161, 230}
  \tikzset{
	  photon/.style={decorate, decoration={snake}, draw=black,very thick},
	  boson/.style={decorate, decoration={snake}, draw=black,very thick},
	  electron/.style={draw=jblue,very thick, postaction={decorate},
	           decoration={markings,mark=at position .55 with {\arrow[draw=black]{>}}}},
	  electron2/.style={draw=jblue,very thick, postaction={decorate},
	           decoration={markings,mark=at position .55 with {\arrow[draw=black]{<}}}},
	  fermion/.style={draw=jblue,very thick, postaction={decorate},
	            decoration={markings,mark=at position .55 with {\arrow[draw=black]{}}}},
	  gluon/.style={decorate, draw=korange,very thick, 
	    decoration={coil,amplitude=4pt, segment length=6pt}},
	  higgs/.style={draw=wred,very thick, postaction={decorate},
	           decoration={markings,mark=at position .55 with {\arrow[draw=black]{>}}}},
	  graviton/.style={draw=wred,very thick, postaction={decorate},decoration={snake}},
	  nothing/.style={draw=white,very thick}
   }
\usepackage[affil-it]{authblk}
\title{Gravity particles from Warped Extra Dimensions, predictions for LHC.}
\author{Alexandra Carvalho \thanks{alexandra.oliveira@cern.ch}}

\affil[1]{Dipartimento di Fisica e Astronomia and INFN, Sezione di Padova, Via Marzolo 8, I-35131 Padova, Italy}
\begin{document}
\pagenumbering{arabic}

\makeatletter
\makeatother
\maketitle

\begin{center}
\section*{Abstract}
\end{center}

Warped Extra Dimension scenarios are a rich playground for phenomenology of heavy resonances at LHC. The Radion and the KK-graviton are inevitable signatures of this class of models. 
On face of the latest LHC experimental results on the direct searches for Beyond Standard Model physics we 
update the phenomenological predictions for production and decay of such resonances in its main production modes.   
We also highlight the phenomenology results relevant to collider searches in individual channels, and provide tools for interpretation of current and future experimental analyses using two specific benchmarks as example.

\clearpage
\setcounter{tocdepth}{2}
\tableofcontents

\section*{Introduction}
\markboth{Introduction}{}%
\addcontentsline{toc}{section}{Introduction}

The richness of LHC results in the search for Beyond the Standard Model (SM) phenomena makes clear that, despite one of main purpose of the LHC experiment to lie in the determination of Higgs boson physics, the search for exotic physics is also a strong branch of the LHC program. The simplest signature of new physics consists in the production of a S-channel resonance that decay to pairs of SM particles. 

Searches for a inclusively s-channel produced heavy particles on LHC data had consistently advanced on both CMS and ATLAS with both 8 TeV and 13 TeV datasets. Several channels were already probed on di-jet \cite{Chatrchyan:2013qha,Khachatryan:2015dcf,CMS-PAS-HIG-16-025,ATLAS-CONF-2016-070,ATLAS-CONF-2016-060,ATLAS-CONF-2016-055,ATLAS:2015nsi}, di-boson \cite{Wang:2014uea,CMS-PAS-EXO-16-034,Khachatryan:2016hje,CMS-PAS-EXO-16-021,Khachatryan:2015cwa,Aaboud:2016trl,Aaboud:2016lwx,Aaboud:2016okv,Aad:2015agg,ATLAS-CONF-2016-082,ATLAS-CONF-2016-062,Aaboud:2016tru}, di-leptons~\cite{Chatrchyan:2012oaa,ATLAS-CONF-2016-061,ATLAS-CONF-2016-059,ATLAS-CONF-2016-045} and di-top~\cite{Kaplan:2008ie,CMS:2014fya,Aad:2012dpa,TheATLAScollaboration:2013kha}. 
Moreover, since the celebrated  discovery of a Higgs boson with mass of $\sim$ 125 GeV~\cite{Chatrchyan:2012ufa,higgsdiscoveryAtlas}
the experimental searches for heavy resonances on LHC are being extended to channels with final states that contain  Higgs particles~\cite{Khachatryan:2016sey,Khachatryan:2015yea,Khachatryan:2015tha,CMS-PAS-HIG-16-029,CMS-PAS-HIG-16-002,Aaboud:2016xco,Aad:2015xja,ATLAS-CONF-2016-049}.  

One point that is not usually in common to all the searches are the physics benchmark used to the searches in the different searches 
and experiments. Even considering the same experimental final state, different benchmarks  may lead to different signal topologies and 
therefore different results. In some cases this discrepancy can make results that should be complimentary not be comparable, as for example the case of searches for pairs of weak bosons (see for example~\cite{Brehmer:2015dan,Dias:2015mhm}). 
In other cases different benchmarks considered are phenomenologically equivalent, as for example the case of the phenomenology of a scalar particle produced by Gluon Fusion (see for example~\cite{Barger:2011hu}). 
 
On this note we focus on the physics of spin-2 and spin-0 resonances. 
Our main purpose is to highlight similarities and differences on the phenomenology of resonances provoked by the structure of its couplings to matter. 
The benchmarks we study are the heavy KK-graviton and radion on the universe scenario of Warped Extra Dimension (WED)~\cite{Randall:1999ee,Goldberger:1999uk}. 

Two scenarios are chosen, the original one where the SM particles are not allowed to propagate along the extra dimension, and the so-called bulk scenario where this constraint is removed. 
We update the predictions for production  and decay using state-of-the-art computation tools for the mentioned resonances to specific model parameters for the following production modes: gluon an quark fusion, Electroweak (EW) associated production with two light jets, associated production with weak bosons and photon fusion. 

Different periods of the LHC program will hold different Center of Mass (CM) energies. 
In this note we derive results for the 8 TeV period of the first run (LHC8) but also to 13 TeV (LHC13) and to 14 TeV (LHC14) as the nominal energy for the machine, energy expected to be reached after the the High Luminosity run (HL-LHC) upgrade. We also produce results to a speculative 100 TeV proton-proton collider. 

The structure of this paper is as follows: 
We remind the motivations of for WED scenarios on section (\ref{sec:WED}). Section (\ref{sec:grav}) introduces the dynamics of the Graviton and Radion fields and Standard Model matter fields. On section (\ref{sec:couplings}) and (\ref{sec:radcoupl}) we present the couplings between the radion and KK-graviton resonances and matter fields. 
Finally in section (\ref{sec:pheno}) we describe the calculations for its phenomenology and discuss the results.  

\section{Warped Extra Dimensions in a nutshell}\label{sec:WED}

\begin{figure*}[hbtp]\begin{center}
\includegraphics[width=0.4\textwidth, angle =0 ]{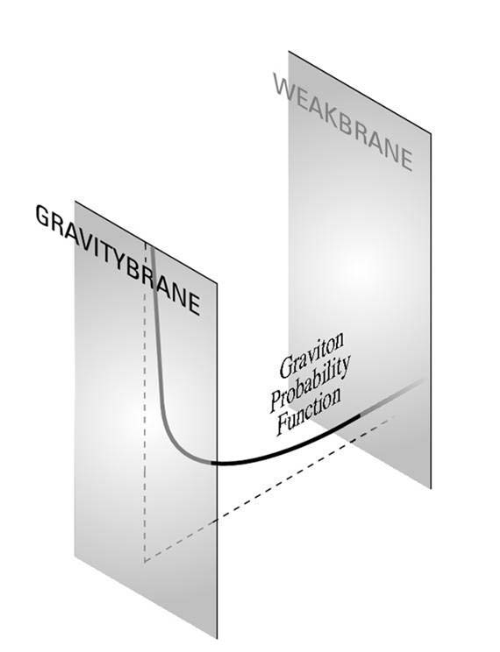}
%
\caption{\footnotesize{Scheme of dimensions on RS theory. The Gravity (Planck) and Weak (TeV) branes are the 4 dimensional boundaries of the extra dimension $\phi$ compactified on a interval ($0,\pi$). The figure also illustrate the metric behavior along the extra dimension. }
\label{fig:ED}}
\end{center}\end{figure*}

A proposition of a non factorizable geometry with one small spatial extra dimension was made on \cite{Randall:1999ee}, the authors exploited a universe scenario with one compact Extra Dimension, the compactification scheme of this dimension allows us to be describe the ED as a line segment between two four dimensional branes, known as Planck and TeV brane (see figure \ref{fig:ED}). 
In this background the most general solution form to solve the classical Einstein motion equations maintaining 4D Poincare\'{e} invariance is:  

\begin{equation}
ds^2 = e^{-2 \sigma(\phi)}\eta_{\mu\nu}dx^{\mu}dx^{\nu}+r_c^2 d\phi ^2\,,
\label{equation:metric}
\end{equation}
where $\mu\nu = 1,...,4$ are the 4D indexes $\phi$ the coordinate for the fifth dimension. 
The full classical action reads:

\begin{equation}
S = S_{Gravity} + S_{TeV} + S_{Planck} + S_{Matter} \,,\label{equation:action}
\end{equation}
where $S_{Gravity}$ is the bulk gravitational action, $S_{Matter}$ is the action matter fields, we consider separately from $S_{Gravity}$ the pieces of gravitational action confined on the branes, we denoted them by $S_{TeV/Planck}$. 
The gravitational part of the action can be written as:

\begin{subequations}
\begin{equation}
S_{i=TeV/Planck} = - \int d^4 x \sqrt{g(\phi=0,\pi)} \Lambda_{i=TeV/Planck} \label{equation:actionSMconf}
\end{equation}
\begin{equation}
S_{Gravity} = \int d^4 x \int_{-\pi}^{\pi} d\phi \sqrt{g} (-\Lambda_{bulk} +2 M_5^3 R)\, \label{equation:actionGr}
\end{equation}
\end{subequations}
where the $\Lambda$'s are vacuum energy densities on the branes/bulk, $R$ is the Ricci tensor, $M_5$ is the 5D Planck mass and $g_{XY}$ is the 5D metric. If we denote the trace of the space-time metric as $g = -g_X^X$ the unity of volume reads $\left(-\int d^5 x \sqrt{g}\right)$. 
Taking $\Lambda_{bulk} = \Lambda_{Planck} = - \Lambda_{TeV} = \Lambda$ we arrive that \cite{Randall:1999ee}:
\begin{equation}
\sigma(\phi) = r_c |\phi|\sqrt{\frac{-\Lambda}{24 M_5^2}} \equiv r_c |\phi| k\,,
\label{equation:compscale}
\end{equation}
the literature refers to the factor $k\equiv\sqrt{\frac{-\Lambda}{24 M_5^2}}$ as the curvature factor, this form strongly depends on the relation between the negative cosmological constant and the brane tensions. Integrating out the extra dimension we find the four dimensional Planck mass to be $\overline{M}_{Pl}^2 = \frac{M_5^3}{k}(1- e^{-2 \pi k r_c})$, where $M_5$ is the five dimensional Plank mass. 
Usually the curvature factor $k$ is assumed to be of the order of 5D Planck scale $M_5$, like this any large value of $k r_c > 1$  does not produce strong hierarchy between the mass constants of the theory ($k,M_5$ and $\overline{M}_{Pl}$). 

The Higgs mechanism can be added to the theory just like in the SM construction\footnote{Althernatives admit the Higgs doublet could be a composite state of the heavy KK fermions, bounded by an additional strong force \cite{Dobrescu:1998dg,Gusynin:2004jp,Hashimoto:2003ym,Bardeen:1989ds}, or even by excitation of QCD \cite{Bai:2008gm}. In the effective theory however the phenomenology for single channels would not differ much from the simple scenario we admit. } with the Higgs mass assumed to be the five dimensional Planck mass $M_5$.   If we confine the Higgs doublet ($H$) to propagate only on the TeV brane its four dimensional vacuum expectation value  is given by  $v \equiv e^{-\pi k r_c}v_0$, taking the 5D Higgs vev $v_0$ of the order of the 5D fundamental mass scale $M_5$ the separation between Planck mass  and EW  scales is produced by the metric when $k r_c \sim 11$. This Planck-EW hierarchy reduction is the most celebrated model feature of Warped Extra Dimensional scenarios. 

Despite this numerical game both $k$ and $r_c$ are free parameters of the theory. 
Probes on the range of validity of classical gravity constrain the value of the compactification radius $r_c$. To do not deal with extremely large/small numbers we use the combinations $\tilde{k} \equiv \frac{k}{\overline{M}_{Pl}}$ and $k r_c$ as the basic parameters of the theory.  
To fix notation and the benchmarks we will consider in this note we review the basic physics and interactions of the gravity particles in the next two sections.

\subsection{Gravity particles}\label{sec:grav}


We refer as {\it gravity particles} the particles resultant of the quantum fluctuations around the classical metric solution (\ref{equation:compscale}). 
The fluctuation modes can be decomposed into a 4D tensor $\otimes$ 4D vector $\otimes$ scalar components:

\begin{equation}
\delta g_{MN}(x,\phi) = \left(\begin{array}{ccc|c}
   &            &  &  \\
   & h_{\mu\nu}(x,\phi) &  & h_{\mu,5}(x,\phi) \\
   &            &  &  \\
  \hline
   & h_{\mu,5}(x,\phi)    & & h_{55}(x,\phi)
 \end{array} \right)
 \label{eq:matrixmetric}
\end{equation}

The motion's relations for the independent fields of (\ref{eq:matrixmetric}) form a over constrained system of equations.
The the axial gauge, defined by fix $h_{\mu 5}=0$, decouples the dynamics of tensor and scalar perturbations. The perturbed space-time metric on on this gauge have the form: 
\begin{equation}
ds^2 = e^{-2 ( \epsilon_r\, h_{55}+\sigma(\phi,r_c))}(\eta_{\mu\nu}+ \epsilon_g h_{\mu\nu}(x,\phi))\,dx^{\mu}dx^{\nu}+r_c^2\, (2\, \epsilon_r\, h_{55}(x,\phi)+1)^2 d\phi ^2 \,.
\label{equation:metricpert}
\end{equation}

To hold control of the order of the expansion we include the expansion parameters $\epsilon_g$ and $\epsilon_r$ on the perturbed metric, with dimensions $[mass]^{-3/2}$ and $[mass]^{-1}$ respectively. 
If we only consider the kinetic terms and the lowest order interaction terms those $\epsilon$'s factorize out of on the particles equations of motion, therefore are not important for Leading order (LO) physics. 
The tensor fluctuations ($h_{\mu\nu}$) correspond to Graviton modes and the scalar fluctuation ($h_{55}$) correspond to Radion mode \cite{Csaki:2000zn,Goldberger:1999un,Charmousis:1999rg}. 
The dynamics and interactions of the gravity particles is found substituting the form (\ref{equation:metricpert}) on the gravitational action $S_{Gravity} = S_{Graviton} + S_{radion}$. 

The five dimensional motion's equations are solved performing a Fourier expansion on the fields five dimensional wave function that separates the $\phi$ and $x_{\mu}$ variables as:
\begin{equation}
X (x^{\mu},\phi)=\sum_{n = 0}^{\infty} X^{(n)} (x^{\mu}) f_X^{(n)}(\phi)\,,\label{eq:kkmodes}
\end{equation}
each four dimensional expansion mode is known as the $n$-th Kaluza-Klein (KK) mode and acts as a particle in the effective four dimensional theory.
The 5th dimensional part of KK-modes wave functions $f_X^{(n)}(\phi)$ are commonly called {\it profiles}. 
The profiles $f_X^{(n)}$ obey a differential Sturm-Liouville problem defined by the equations of motion, spin of the field and boundary conditions, the profiles for all the known fields are described by a combination of exponential and Bessel functions. A didactic guide to the behavior of matter fields on bulk WED scenario can be found on~\cite{Gherghetta:2010cj}. It is out of scope of this note to write down and solve the 5D equations of motion and derive in details each matter field 5D profile, we just highlight the results necessary for further discussion. 

We denote the Graviton four dimensional wave functions (profiles) as $h_{\mu\nu}^{(n)}(x_{\mu})$  ($f_h^{(n)}(\phi)$). The zero mode of the Graviton field correspond to the mediator of gravitational force.  The first graviton KK-mode therefore have a profile that is TeV localized, and its 4D effective mass is:
\begin{eqnarray}
m_{Gr} \equiv m_{Gr}^{(1)} = x_1 \tilde{k} e^{-  k L} \overline{M}_{Pl} \sim \mbox{TeV}\,,
\end{eqnarray}
where $x_1 = 3.83$ is the first zero of the Bessel function $J_1$. If we want to use as the free parameter of theory  $m_{Gr}$ and $\tilde{k}$ we must use the relation:
\begin{equation}
k r_c = \frac{1}{\pi} Ln\left( x_1 \tilde{k} \frac{\overline{M}_{Pl}}{M_{Gr}} \right)\,.
\label{eq:paramspace}
\end{equation}
The dependence of $k r_c$ on the parameters $M_{Gr}$ and $\tilde{k}$ is very mild. Following the above relation the $k r_c$ value varies between 10 and 12 if we profile $\tilde{k}$ vary between 0.01 and 1 and $M_{Gr}$ between 100 GeV and 1.5 TeV. Remaining a phenomenologically acceptable value to reproduce the mass hierarchy between the Planck mass and the electroweak scale on a wide range of conditions. 
 
A similar procedure defines the physics of the scalar fluctuation of the metric, known as Radion. 
To the five dimensional Radion field $r(x,\phi)$ have canonical kinetic term, this field relates to $h_{55}(x,\phi)$ as~\cite{Dominici:2002jv}: 
\begin{equation}
r(x,\phi) = \sqrt{\frac{24 M_5^3}{k}}\,e^{-k\, \epsilon_r h_{55}(x,\phi)\pi}\,.
\end{equation} 
If we consider terms with only two insertions of the $r$ field, $\epsilon_r$ is an irrelevant parameter to the theory . 

In the simplest case the zero mode of the Radion (that we will simply call radion)  would behave as a massless field~\cite{Charmousis:1999rg}. The radion however couples with the SM particles trough expansions of the metric on $S_{Matter}$ action, including photons~\cite{Maiani:1986md}, slightly modifying propagation of light, what is experimentally disfavored~\cite{Dobrich:2010hi,Tam:2011kw}.
In addition, the distance between branes ($r_c$) was introduced in the un-perturbed metric~(\ref{equation:metric}) as a free parameter, even thought its value is an important feature on the theory. The value of $r_c$ is important to define the hierarchy of gravity and weak scales meaning it is desirable to exist one mechanism that dynamically fix its value. 

Reference \cite{Goldberger:1999uk} proposed as solution that the compactification radius could be dynamically generated as vacuum expectation value of an additional bulk scalar $\Phi$ with a potential $V(\Phi)$. The potential energy would be created by the existence of a bulk scalar field with specific brane potential~\footnote{ The references~\cite{DeWolfe:1999cp,Medina:2010mu} also worked out examples in terms of benchmark super-potentials.  This generic idea can be adapted to specific models, for example \cite{Luty:1999cz} suggested the Radion potential could be stabilized by the condensation of bulk super-symmetric particles.}. 
As a result of the back-reaction of $\Phi$ in the Lagrangian the radion acquire a small mass term, dependent of the fundamental parameters of its potential. The dynamically generated radion mass $m_r$ is expected to be small in comparison with the mass scale of the first KK-modes (for example $m_{Gr}$). We assume $m_r$ as an theory input parameter. 

In general, for a given $V(\Phi)$ the metric solution and the value for the stabilized $r_c$ have to be calculated numerically. 
However regardless the functional form of the class of  potentials postulated in ~\cite{Goldberger:1999uk} the extra dimensional profile function $f_r^{(0)}(\phi)$ is TeV localized by the factor $e^{2 k (\phi +\pi) r_c}$. We can extend this conclusion to any general form of $V(\Phi)$. 

\section{KK-graviton couplings to matter}\label{sec:couplings}

We consider two main benchmarks for the couplings of the gravity particles with matter fields, those are defined by how the SM matter fields behave in the extra dimension: 
On the original work the SM matter is localized on the TeV brane, as the Higgs doublet. The possibility of matter fields to propagate in the bulk of the extra dimension under the space-time started to be investigated few time after their publication (see for example~\cite{Goldberger:1999wh,Gherghetta:2000qt}). 
We will refer to the first as RS1 model and the second as bulk model. 

\begin{figure*}[h]\begin{center}
\includegraphics[width=0.48\textwidth, angle =0 ]{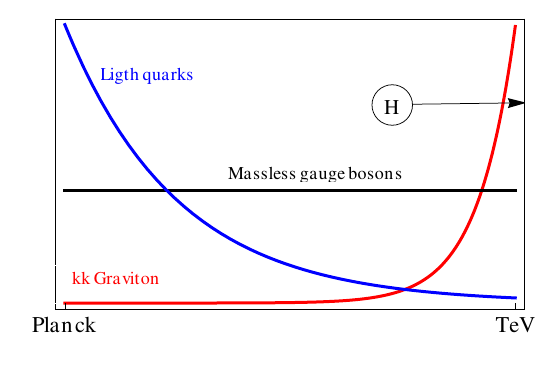}
\includegraphics[width=0.48\textwidth, angle =0 ]{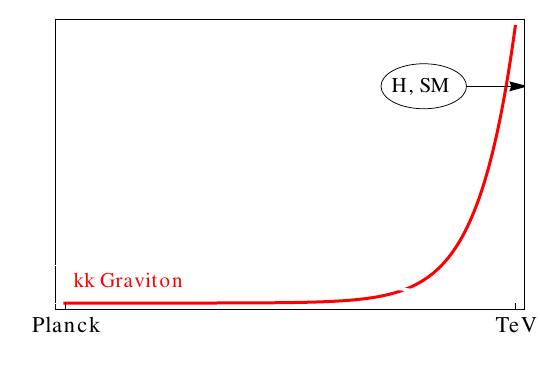}
\caption{\small{Scheme of matter localization on the different WED scenarios. {\bf Left:} Bulk scenario. {\bf Right:} RS1 scenario. The combination of Exponential and Bessel functions makes the KK-graviton profile to be very TeV localized. To do not overload the figure we do not show the profile of zero mode Graviton, neither of the third generation of fermions. }\label{figure:bulkWED}}
\end{center}\end{figure*}

In the bulk scenario a KK expansion similar to the one in equation~\ref{eq:kkmodes} is applied to each field. 
The zero mode of each KK tower correspond to the SM correspondent SM particle, and assumed to be originally massless. 
The derivation of the functional form of the profiles for each field hypothesis can be found for example in~\cite{Gherghetta:2010cj}. 
In figure (\ref{figure:bulkWED}) we draw a schematic view of the profiles of the zero modes matter fields along the extra dimension $\phi$ in contrast to the KK-graviton. 

The effective four dimensional strength of the interaction between any fields are proportional to the integral of their profiles in the fifth dimension and 
the interactions of the KK-graviton with matter are suppressed by $\Lambda_G \equiv \frac{m_G}{x_1 \tilde{k} }$. 
This results in the following Lagrangian: 
\begin{equation}
{\cal L} =  - \frac{x_1 \tilde{k} }{m_G}  h^{\mu \nu (1)}\times d_i T_{\mu \nu}^{i}\,, 
\label{eq:LG}
\end{equation}
where $T_{\mu \nu}^{i} \equiv - \frac{2}{\sqrt{g}} \frac{\delta L}{\delta g^{\mu \nu}}$ is the four dimensional canonical energy-momentum tensor for the field $i$ \cite{Forger:2003ut} and $d_i \equiv \int d\phi f_{Gr} f_i^2$ is the integral overlap between the profiles of the fields $i$ with KK-graviton.  
On bulk scenario all the profile functions are normalized, the parameter $d_i$ can only be one or less.
Being the $f_{Gr}$ extremely TeV localized the parameter $d_i$ indicates how much the profile of the field $i$ is away from TeV brane. 
For example, if the Higgs doublet is confined to TeV brane the volume suppression parameter in the coupling strength is:
\begin{equation}
d_H = f_h(\phi =\pi) \sim 1\,.
\end{equation}
On the  RS1 scenario all the particles are TeV localized therefore the strength of the couplings between KK-graviton and SM matter are democratic between each field degree of freedom.   
In the following sections we derive the KK-graviton coupling strengths to gauge bosons in terms of the volume suppression terms~$d_i$. 

\subsection{Massless gauge bosons}

\begin{figure*}[hbtp]\begin{center}
\includegraphics[width=0.40\textwidth, angle =0 ]{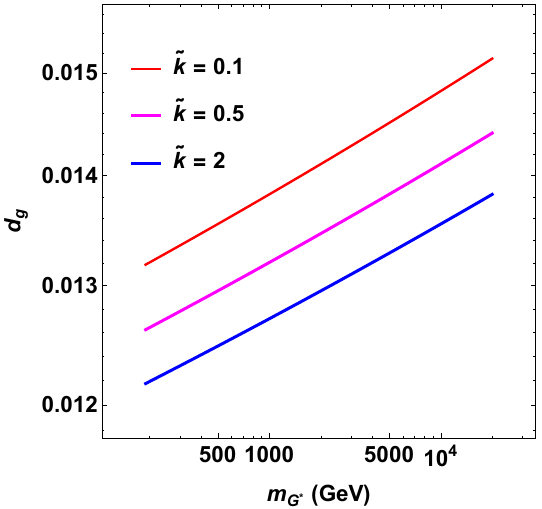}
\caption{\footnotesize{The $m_{G^*}$ dependence of volume suppression in the couplings between  $G^*$ and massless gauge bosons in the bulk scenario for different values of $\tilde{k}$.  
 \label{figure:bulkWEDcg}}}
\end{center}\end{figure*}

We understand as massless gauge bosons gluons, photons and the transverse vector bosons $V = W, Z$ in the format of their field strength. 
As outlined on figure (\ref{figure:bulkWED}) the profile of a massless zero mode of a spin-1 bulk boson is flat with respect to the KK-graviton profile. 
The volume suppression in the four dimensional KK-graviton coupling to massless gauge bosons by the factor:
\begin{equation}
d_{g} = \frac{2}{k \pi r_c } \frac{ (1-J_0 (x_1))}{x_1^2 |J_2(x_1)|}
\label{eq:cg}
\end{equation}
For illustration purposes in figure (\ref{figure:bulkWEDcg}) we draw the dependence of the $d_g$ parameter with the graviton mass $M_{Gr}$, fixing the dimensionless parameter $\tilde{k}=$0.5, 1 and 2. The ratio $d_{g}(\tilde{k}=0.1)/d_{g}(\tilde{k}=2)$ is $\sim$8\%-9\%.  

\subsection{Weak bosons}

The W and Z masses are proportional to the vacuum expectation value of the Higgs doublet after EWSB. 
Their longitudinal degrees of freedom in the energy momentum tensor result from absorbing the Goldstone bosons of the Higgs doublet. 
The couplings of the KK-graviton with the massive vector bosons in the unitary gauge can be understudied in two components: a component related with their energy momentum tensor 
(volume suppressed by the $d_g$ parameter, as a massless gauge boson) and the component that arises from EWSB that feels no volume suppression. 
The coupling of the KK-graviton to a pair of massive vector bosons is given by:
\begin{equation}
d_V T^{VV}_{\mu\nu} =  
-  \left( m_W^2 \frac{\delta(g_{\alpha\beta}W_{+}^{\alpha}W^{\beta}_{-}) }{\delta g^{\mu\nu}}  + \frac{m_Z^2}{2} \frac{\delta(g_{\alpha\beta}Z^{\alpha}Z^{\beta}) }{\delta g^{\mu\nu}}  \right)
- d_g  T_{\mu\nu}^{YM}(W^{\pm},Z,\gamma) ,
\label{eq:GVV}
\end{equation}
where  $T_{\mu\nu}^{YM}(W^{\pm},Z,\gamma) = g_{\alpha\beta}\frac{1}{4}F^A_{\mu\nu} F_{A,\mu\nu} + F^{A,\mu}_{\alpha} F^{A,\beta\mu}$ is the canonical Yang Mills energy-momentum tensor for $SU(2)_L \times U(1)_Y$ of gauge fields. 

\subsection{Fermions}

Similarly to the last section we can understand the coupling of KK-graviton with the physical fermions as separated in tree parts: two parts devoted to the (massless) Dirac energy momentum tensor of each chirality $T^{\psi_{L/R}}_{\mu\nu} = \overline{\psi}_{L/R} \slashed{\partial} \psi_{L/R}$, and other part proportional to the EWSB mass:
\begin{equation}
d_{\psi} T^{\psi}_{\mu\nu} =  d_{f_L} T^{\psi_L}_{\mu\nu} + d_{f_R} T^{\psi_R}_{\mu\nu} 
+ d_{f_{LR}} g_{\mu\nu} \lambda_f v^2 (\bar{\psi_R} \psi_L + h.c.)\,, \label{eq:gravferm}
\end{equation}
where $\psi$ is the four dimensional fermion field and $d_{LR} \equiv \int f_{Gr} f_{f_R} f_{f_L}$. 
In the case of the fermions there is a partial freedom to choose the profiles localization of different fermion chiralities, that is propagated as freedom to fix KK-graviton couplings to bulk fermions.

This freedom is only partial, since the four dimensional mass of the fermions is directly related with their profiles localization's (see for example \cite{Gherghetta:2000qt}). 
To fix the mass of the light fermions we need to choose both fermion chiralities to be Planck localized, resulting the constants $d_{f_L} = d_{f_R} = d_{f_{LR}} \sim 0$.  We can ignore the KK-graviton couplings to bulk light fermions, namely leptons and light quarks.  

The choice of fermion localization parameters is a delicate issue when considering the third family of quarks and the related experimental constraints. 
 For practical reasons we consider as benchmark the case where only the SM gauge symmetry $SU(2)_L\times U(1)_Y$ is enlarged to the bulk and 
$d_{f L} \sim d_{f LR} \sim 0$ and $d_{f R} \sim 1$, we will refer to this choice as consider a elementary top quark case~\cite{Fitzpatrick:2007qr}. 

Tree geometrical parameters controls the profile localization of each generation $c_{3_L}$ (for the left handed doublet) and $c_{t_R}$ and $c_{b_R}$ (for the right handed singlets). The $t_L$ and $b_L$ are zero modes of a bulk $SU(2)_L$ doublet and share a common profile.
To reproduce both the top and bottom quark masses (with two orders of magnitude distance) we need one of the $L/R$ profiles to be TeV localized. 
The simplest solution for the third family embedding on the WED bulk (on the elementary top scenario) is to choose the $t_R$ profile to be TeV localized and the $t_L$ profile to be Planck localized ($c_{t_R} \to 1$ and $c_{t_L} \to 0$)~\cite{Hewett:2002fe}. 
As a result the presence of heavy KK modes produces dimension 6 operators in the low energy theory produces corrections in the SM predictions by inducing high dimension operators. The pure bosonic operators contribute at tree level to the oblique parameters $S$ and $T$~\cite{Peskin:1991sw,PhysRevLett.65.2967}, the fermionic operators contribute non-obliquely correcting processes like $Z\to b_L\bar{b}_L$~\cite{Agashe:2003zs} .  This scenario leads to very stringent limits on the KK-fermion masses ($O(10)$ TeV), and on the masses of the KK-gauge-bosons  ($O(20)$ TeV) \cite{Csaki:1999mp,Davoudiasl:2000wi}. 

The authors of references \cite{Agashe:2003zs,Carena:2006bn} realized that enlarge the gauge symmetry on the WED bulk to a $SU(2)_L\times SU(2)_R$ custodial symmetry causes a partial cancellation of the beyond tree level corrections caused by the heavy KK-modes on electroweak corrections from both gauge \cite{Agashe:2003zs} and fermion sector \cite{Carena:2006bn}, lowering down the indirect limits on the mass scale of matter KK modes. 
To enlarge the SM custodial Symmetry to the WED bulk however opens the possibility for the existence new light fermions with exotic charge \cite{Carena:2006bn}. 
It is not the scope of this document to discuss the model building hypothesis inside the microscopic WED scenario. 
We exemplify the deviations KK-graviton phenomenology can suffer due model building on fermion sector by comparing the total width and decay rates of the scenario \cite{Fitzpatrick:2007qr} with the extreme hypothesis of consider the profiles of both top chiralities to be Planck localized considering an additional benchmark where $d_{top} = 0$.

The covariant derivative in the fermion energy momentum tensor ($T^{\psi_{L/R}}_{\mu\nu}$) induce an additional interaction term of SM fields with the KK-graviton with the form~\cite{Giudice:1998ck,Han:1998sg}:
\begin{equation}
d_{fV}\, T^{5}_{\mu\nu} = d_{fV}\, eQ  \bar{\psi_R} ( \gamma_{\mu} A_{\nu} +\gamma_{\nu} A_{\mu} ) \psi_L + h.c.\,,
\label{eq:AGff}
\end{equation}
where $eQ = \frac{\alpha_S}{4 \pi}$ for the gluon, $eQ = \frac{\alpha_{EW}}{4 \pi}$ for the foton and  $eQ = \frac{\alpha_{EW}}{4 \pi\, \sin(\theta_W)} $ $\left( eQ = \frac{\alpha_{EW}}{4 \pi\, \cos(\theta_W)}\right) $ for the $W$ ($Z$) boson. 
These interactions are relevant for electroweak processes as the of a $G^*$ in association with a vector or in association with two jets. 
As already pointed out, in bulk scenario any interaction terms of the KK-graviton are suppressed by the overlap between the five dimensional profiles of the interacting particles. 

\section{Radion couplings to matter}\label{sec:radcoupl}

The four dimensional couplings of the radion with matter are also derived performing the KK expansion of the radion field in the full Lagrangian. 
In this section we fix the notation for the normalization of the radion couplings to matter for the different  benchmark scenarios following the notation of the references~\cite{Csaki:2002gy,Rizzo:2002pq,Cheung:2000rw}. 

The radion couplings are proportional to the trace of the EMT following the Lagrangian:
\begin{equation}
{\cal L} = - \frac{r}{\Lambda_R}  \times a_i T_{\mu}^{\mu (i)}(x)\,,   
\label{eq:LR}
\end{equation}
where $\Lambda_R \equiv \sqrt{6} \times k e^{- k \pi r_c} \sqrt{\frac{M_5^3}{k^3}}$ is treated as an model parameter. 
The zero-th order energy-momentum trace of massive particles proportional to their masses.   
We neglect the possibility of brane localized terms for the gauge fields that could contribute with the tree level coupling.

In the RS1 scenario of the coupling between the radion and massive gauge bosons is given by:
\begin{equation}
a_V T_{\mu}^{\mu (V)}(x) = \left(\int d\phi f_H^2 f_r \right)\times 2 (m_W^2 W_{+\mu}W^{\mu}_{-}   + \frac{m_Z^2}{2} Z_{\mu}Z^{\mu} ) \,.
\end{equation}
The couplings of the radion to each fermion $f_i$ are:
\begin{equation}
a_{f_i} T_{\mu}^{\mu (f_i)}(x) = \left(\int d\phi f_H^2 f_r  \right) 
\times m_{f_i}  (\overline{\psi}_{f_iL}\psi_{f_iR} + h.c.)\,.
\end{equation}

As we assume the profile of the radion field to be TeV localized (see section~\ref{sec:WED})  the strength of the couplings between the radion and a massive SM states is: 
\begin{equation}
\int d\phi f_H^2 f_r \sim 1
\label{eq:ah}
\end{equation}

In the case of massless gauge bosons non-vanishing part of the trace of EMT arises through loop level trace anomalies, namely the $\beta$ function of is coupling constant. 
The LO coupling of a radion couplings to both gluon and photon in the RS1 scenario are: 
\begin{equation}
a_{g} = \frac{1}{4} \frac{\alpha_{s}}{2\pi} \beta_{QCD}  \,\,\,\,\, ,
 \,\,\,\,\,\, a_{\gamma} = \frac{1}{4} \frac{\alpha_{QED}}{2\pi} \beta_{QED} \, ,
\end{equation}
Radiative corrections in both QCD and QED are important ingredients to the three level radion phenomenology. 
We use as reference one loop beta functions and the the two loop evolution to $\alpha_{s}$~\cite{book}. 

In the bulk scenario the expansion of the bulk field strength $F_{MN}$ on space-time components, induces the action:
\begin{equation}
S_{bulk}^{(g,\gamma)} \equiv - \frac{1}{k r_c\, \pi} \int d^4 x \quad  
\frac{r}{4 \Lambda_R}  
\left(
 \sum_{colors} T_{\mu\nu}^{QCD} T^{\mu\nu, QCD}+
T_{\mu\nu}^{EW} T^{\mu\nu, EW}
\right)\,,
\label{eq:bulktermglu}
\end{equation}
that generates a bulk term contributions to the radion couplings strengths. 
Finally, the LO coupling of a radion couplings to both gluon and photon in the bulk scenario are: 
\begin{equation}
a_{g} = \frac{1}{ 4}  \left( \frac{\alpha_{s}}{2\pi} \beta_{QCD} + \frac{1}{ k \pi r_c} \right) \,\,\,\,\, ,
 \,\,\,\,\,\, a_{\gamma} = \frac{1}{4}  \left(  \frac{\alpha_{QED}}{2\pi} \beta_{QCD} + \frac{1}{k r_c\,\pi} \right)\, ,
 \label{eq:rggbulk}
\end{equation} 
while the interaction between the radion and massive gauge bosons is: 
\begin{equation}
a_V T_{\mu}^{\mu (V)}(x) = 2 (\mu_W^2 W_{+\mu}W^{\mu}_{-}   + \frac{\mu_Z^2}{2} Z_{\mu}Z^{\mu} )
+ \frac{1}{4 k r_c\,\pi} T_{\mu\nu}^{EW} T^{\mu\nu,  EW}\,,
\label{eq:rvvbulk}
\end{equation}
where $\mu_i^2 = m_i^2 \left( 1- \frac{1}{k \pi r_c} \frac{m_i^2}{(k e^{-k\pi r_c})^2} \right)$, to $i = W, Z$~\cite{Csaki:2002gy}. 
Effectively, $\mu_i$ represent the physical masses of the W and Z bosons.  
In the rest of the document we ignore the bulk contribution to $\mu_i$ and assume $\mu_i^2 = m_i^2$.

\section{Phenomenology at LHC}\label{sec:pheno}

From now on we will denote the KK-graviton as $G^*$ and to the radion as $R$, or collectively by $X$. 
We study the production of a  $X$ particle in proton-proton collider induced by QCD induced and EW processes to $m_{X}$ ranges between 200 GeV and $E_{CM}/2$, where $E_(CM)$ is the energy of the collision. 
We present calculations of cross sections, total width and branching ratios in the benchmarks reviewed last section. 

The QCD induced production modes we consider are gluon fusion (GF) and quark fusion (QF). 
The EW induced processes are photon fusion (PF) and the associated production with two prompt jets (Xjj). 
This last is partially composed by Vector Boson Fusion (VBF) and by the associated production with a hadronic massive vector boson (VX).
When relevant, the results are provided for $\tilde{k} = 0.1$, $k r_c \pi= 35$ and $\Lambda_R =3$~TeV. 
It is to easy to re-scale the results to any other choice of model parameters using analytic formulas, in summarized in appendix~\ref{app:tables}.  

Four Center of Mass (CM) energies for proton-proton collisions are considered: 8 TeV, 13 TeV, 14 TeV and 100 TeV that we will denote 
respectively LHC8, LHC13, LHC14 and LHC100. 
The first correspond to the energy of the Run 1 of LHC and the second to the Run 2. 
The plans for the LHC is to arrive to 14 TeV in the HL-LHC phase around 2022, and eventually a 100 TeV proton-proton machine may also be constructed in a rainbow future. For brevity we only plot results for 13 TeV collisions, the results for the other CM energies are shown in appendix~\ref{app:tables}. All the results are available in electronic form here~\cite{WEDgit}.

If not mentioned we calculate cross sections  with {\tt MG5\_aMC@NLO}~\cite{Alwall:2014hca} interfaced to {\it LHAPDF6}~\cite{Buckley:2014ana}.
We use  PDF4LHC15\_nlo\_mc\_pdfas set~\cite{Carrazza:2015hva, Butterworth:2015oua, Dulat:2015mca,Harland-Lang:2014zoa, Ball:2014uwa} and the four flavor scheme for the proton. The central value for the strong coupling is taken as $\alpha_s(m_Z) = 0.118$. The $\alpha_S$ scale uncertainty is calculated two extra replicas are calculated $\alpha_s(m_Z) = 0.1165$ and $\alpha_s(m_Z) = 0.1195$ to define the corresponding uncertainty $\delta_{\alpha_S}$.  
The renormalizarion and factorization QCD scales are taken as $\mu_R =  \mu_F = \mu = m_{Gr}$. For the VBF and VX production modes we use floating scales.  
The relative uncertainty is estimated in the range $1/2 < \mu/M_{Gr} <2$, while the $\mu_F/$ and  $\mu_R$ correlation is taken from fully correlated up to 10\% correlated in steps of 10\%. The scale uncertainty $\delta_{scale}$ is estimated by the root mean square of those calculations and the full uncertainty is the square sum of $\delta_{PDF}$, $\delta_{\alpha_S}$ and $\delta_{scale}$. 

The inclusive photon fusion production is calculated using the QED NNPDF23@NNLO ($\alpha_S$=0119) MC set with 100 replicas~\cite{Ball:2013hta}, with the same definition of scales and related uncertainties used for the GF production. 
To each data point we generate 20,000 events, makings the MC statistical uncertainty is negligible.
We fix $\Gamma_{X} = 1$~GeV independent of the resonance mass. The effect of finite width in the total cross section is subleading in the region narrow width approximation is valid. 

\subsection{KK-graviton production and decay}\label{sec:numbers}

To calculate the QCD induced $G^*$ production cross section we use a model implementation capable to calculate Next to Leading Order (NLO) QCD corrections to the GF production of a spin-2 particle~\cite{Artoisenet:2013puc}\footnote{In this implemention the couplings of the $G^*$ to light quarks and gluons  are independent input parameters ($k_q$, $k_g$), together with the 
theory cuttoff ($\Lambda$). The relation among these parameters and the ones considered in this document is $\frac{k_g}{\Lambda} = \frac{x_1 \tilde{k}}{m_{G^*}}$. To avoid numeric UV divergences in the NLO calculation one cannot consider $k_q$ initially zero, therefore when considering the bulk scenario we assume a small $k_q = 5\times 10^{-3}$, requesting calculation precision of 1\%. }. The LO results obtained with this models are checked against the ones from the model implemented by~\cite{deAquino:2011ix, aquino} from {\tt Feynrules}~\cite{Christensen:2008py} database in {\tt MG5\_aMC@NLO}~\cite{Alwall:2014hca}, as well with the WED implementation default on {\tt Pythia6} \cite{Pythia6-0}. As a further cross check for the  bulk scenario we modify the model~\cite{deAquino:2011ix, aquino}, introducing  the coupling modifications highlighted in section~\ref{sec:couplings}, this model can be found in~\cite{WEDgit} and its found to agree with the bulk WED scenario implemented by the authors of~\cite{Tuomas} on {\tt CalcHep}~\cite{Belyaev:2012qa} framework and with the analytic formulas for $G^*$ branching ratios presented here.

\begin{figure*}[h]\begin{center}
\includegraphics[width=0.44\textwidth, angle =0 ]{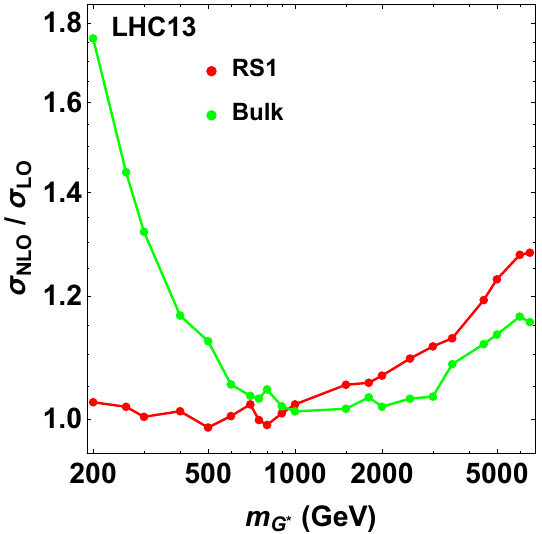}
\includegraphics[width=0.44\textwidth, angle =0 ]{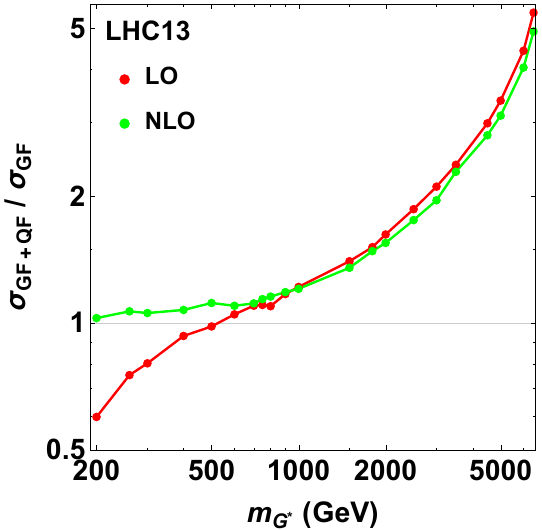}
\caption{\small{
{\bf Left:} K-factors for QCD induced production cross section of a $G^*$ for a 13 TeV collider.  
{\bf Right:} Ratio between a pure gluon fusion and a pure quark fusion production of a spin-2 particle  in the RS1 scenario  at LO and at NLO in QCD.  
 }\label{figure:kfac}}
\end{center}\end{figure*}

In the RS1 scenario a $G^*$ is produced by both QF and GF modes.   
The production mode make an effect in the $G^*$ kinematics: At LO the quark initiated process is more asymmetric in the plane of the collision lowering the $G^*$ transverse momenta.  
Also, at NLO level the production modes interfere, impacting in the total cross section.  
In figure~\ref{figure:kfac} we show the k-factors for the QCD induced production and the ratio between the different components of the QCD induced $G^*$ in the RS1 scenario at LO and NLO in QCD. 
From the left figure we see effect of the destructive interference between the diagrams with quark and gluon coupling to $G^*$ when $m_{G^*} <  700$. This effect makes the k-factor of the RS1 case to be negligible. When $m_{G^*} > \sim$2 TeV the RS1 cross section is dominated by QF and the k-factors  can arrive to 25\%. 

\begin{figure}[h]
\centering
\begin{tabular}{ ccc } 
		\begin{tikzpicture}[
				thick,
				level/.style={level distance=2cm, line width=0.4mm},
		  		level 2/.style={sibling distance=1.5cm},
		  		level 3/.style={sibling distance=1.5cm}
			]
			\coordinate
			child[grow=right] 
			{
				child [grow=down, level distance=0.9cm] {
		      		child[grow=right] {
				 		edge from parent [graviton]	
				 		node [above] {$G*$}	
		       	}
		       		child[grow=right] {
				 		edge from parent [boson]		
		       	}
		      		child[grow=down, level distance=0.9cm] {
		         		child[grow=left] {
		            		edge from parent [electron]	
		              	node[below] {$q'$}
		            }	
		           	child[grow=right] {
		              edge from parent [electron2]	
		              node[below] {$q'''$}
		            }	
		            edge from parent [boson]	
					}
					edge from parent [boson]
				}
				child[grow=east] {
					edge from parent [electron]
					node [above] {$q''$}
				}
				edge from parent [electron]
				node [above] {$q$}
			};
		\end{tikzpicture} &
		\begin{tikzpicture}[
				thick,
				level/.style={level distance=2cm, line width=0.4mm},
		  		level 2/.style={sibling distance=1.5cm},
		  		level 3/.style={sibling distance=1.5cm}
			]
			\coordinate
			child[grow=right] 
			{
		      		child[grow=down, level distance=1.3cm] {
		         		child[grow=left] {
		            		edge from parent [electron]	
		              	node[below] {$q$}
		            }	
		           	child[grow=right] {
		              edge from parent [electron2]	
		              node[below] {$q''$}
		            }	
		            edge from parent [boson]	
					}
					child[grow=-10] {
		            edge from parent [graviton, solid]
		            node [below=3pt]{$G^*$}
					}
					child[grow=-10] {
		            edge from parent [boson, solid]
					}
					child[grow=east] {
						node [above]{$q'''$}
		            edge from parent [electron]
					}
				edge from parent [electron]
				node [above] {$q'$}
			};
		\end{tikzpicture} &
		\begin{tikzpicture}[
					thick,
			level/.style={level distance=1.4cm, line width=0.4mm},
			level 2/.style={sibling distance=1.4cm},
			level 3/.style={sibling distance=1.4cm},
			level 4/.style={sibling distance=1.4cm}
	]
			\coordinate
			child[grow=-30]{
				child[grow=east] { 
							child[grow=-40] {
								edge from parent [graviton]
		              		   node[below] {$G^*\qquad$}
							}
							child[grow=-40] {
								edge from parent [boson]
							}
                            child[grow=0] {
                               node[above] {$q'''$}
								edge from parent [electron]
							}
					child[grow=40] {
					   node[above] {$q''$}
						edge from parent [electron2]
					}    
		  			edge from parent [boson]      
		 		}
				child[grow=210] {
		   			edge from parent [electron]
		   			node[below=3pt] {$q$}
		 		}
		 		edge from parent [electron] node [above=3pt] {$q'$}
			};
		\end{tikzpicture} \\
(a) & (b) & (c) \\
\end{tabular}
\begin{tabular}{ ccc }
		\begin{tikzpicture}[
					thick,
			level/.style={level distance=1.4cm, line width=0.4mm},
			level 2/.style={sibling distance=1.4cm},
			level 3/.style={sibling distance=1.4cm},
			level 4/.style={sibling distance=1.4cm}
	]
			\coordinate
			child[grow=-30]{
				child[grow=east] { 
							child[grow=-40] {
								edge from parent [graviton]
		              		   node[below] {$G^*\qquad$}
							}
							child[grow=-40] {
								edge from parent [boson]
							}					    
					child[grow=40] {
						edge from parent [boson]
						node [above] {$V\qquad$}
					}    
		  			edge from parent [boson]
		 		}
				child[grow=210] {
		   			edge from parent [electron]
		   			node[below=3pt] {$q'$}
		 		}
		 		edge from parent [electron] node [above=3pt] {$q$}
			};
		\end{tikzpicture} &
		\begin{tikzpicture}[
				thick,
				level/.style={level distance=2cm, line width=0.4mm},
		  		level 2/.style={sibling distance=1.5cm},
		  		level 3/.style={sibling distance=1.5cm}
			]
			\coordinate
			child[grow=right] 
			{
		      		child[grow=down, level distance=1.3cm] {
		         		child[grow=left] {
		            		edge from parent [electron]	
		              	node[below] {$q$}
		            }	
		           	child[grow=right] {
		              edge from parent [graviton]	
		              node[below=3pt] {$G^*$}
		            }	
		            child[grow=right] {
		              edge from parent [boson]	
		            }	
		            edge from parent [electron]	
					}
					child[grow=right] {
		            edge from parent [boson]
		            node [above=3pt]{$V$}
					}
				edge from parent [electron]
				node [above] {$q'$}
			};
		\end{tikzpicture} &
		\begin{tikzpicture}[
					thick,
			level/.style={level distance=1.4cm, line width=0.4mm},
			level 2/.style={sibling distance=1.4cm},
			level 3/.style={sibling distance=1.4cm},
			level 4/.style={sibling distance=1.4cm}
	]
			\coordinate
			child[grow=-30]{
							child[grow=-40] {
								edge from parent [graviton]
		              		   node[below] {$G^*\qquad$}
							}
							child[grow=-40] {
								edge from parent [boson]
							}
					child[grow=40] {
		  			edge from parent [boson]
		   			node[above=3pt] {$V\,\,\,$}           
		 		}
				child[grow=210] {
		   			edge from parent [electron]
		   			node[below=3pt] {$q$}
		 		}
		 		edge from parent [electron] node [above=3pt] {$q'$}
			};
		\end{tikzpicture} \\
(d) & (e) & (f) \\
\end{tabular}
\caption{\small{
Feynman diagrams for EW induced $G^*$ production. 
The processes (a), (b) and (c) stand for production in association with two prompt jets 
while (d), (e) and (f) are for associated production with an massive vector boson. 
The processes (b), (c), (e) and (f) are only present in the RS1 scenario. 
}}
\label{fig:resonanthh}
\end{figure}
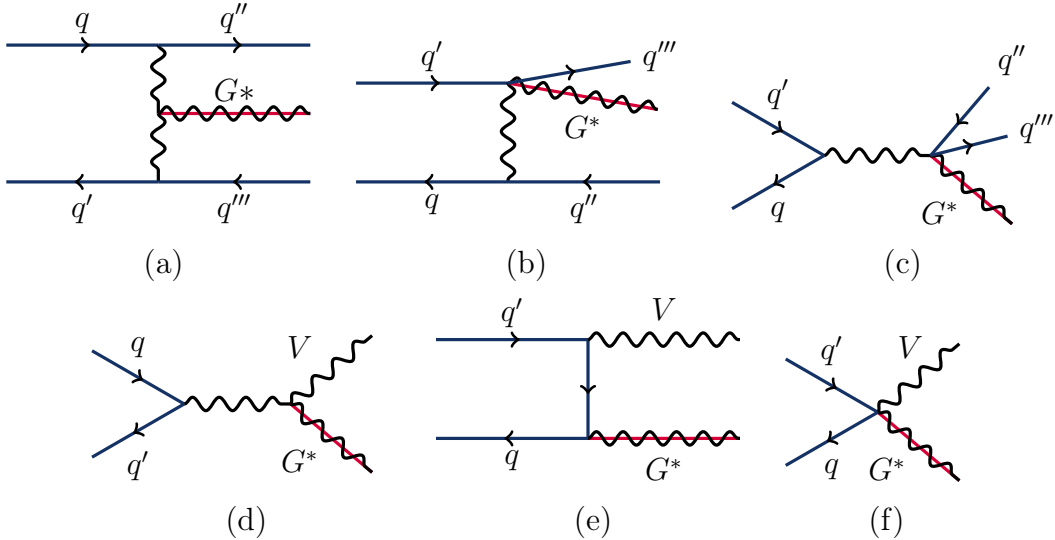

The EW induced associated production with two jets sketched in figure \ref{fig:resonanthh}, several subprocesses compose the signal: 
 the processes (a) is the so-called Vector Boson Fusion (VBF) production, the (b) and (c) are induced by the dimension-5 operator discussed in section \ref{sec:radcoupl}; the associated production with a massive vector boson $V = W, Z$ when it decays hadronicaly is also included (d, e and f). 
 In the bulk scenario only (a) and (d) are non negligible. 
We had used the model~\cite{WEDtwiki} to simulate the bulk scenario and~\cite{deAquino:2011ix, aquino} to the RS1 scenario. 
 The process (e) can be interpreted as the EW correction of the QCD induced processes, and  is negligible in the hole $m_{Gr}$ range. 
 
The (simbolic) generation level selection applied on di-jet system are $p_{t,j} > 1 \; \mbox{GeV}$, $\Delta R_{jj} > 0.1$ and $\eta_j >7$, where $p_{t,j}$ is the transverse momenta of each quark in the final state, $\Delta R_{jj}$ is the angular distance between the hard jets and $\eta_j$ the jet pseudo-rapidity.  The loose jet selection assures that all the signal cross section is considered in inclusive searches.

\begin{figure}[h]\begin{center}
\includegraphics[width=0.45\textwidth, angle =0 ]{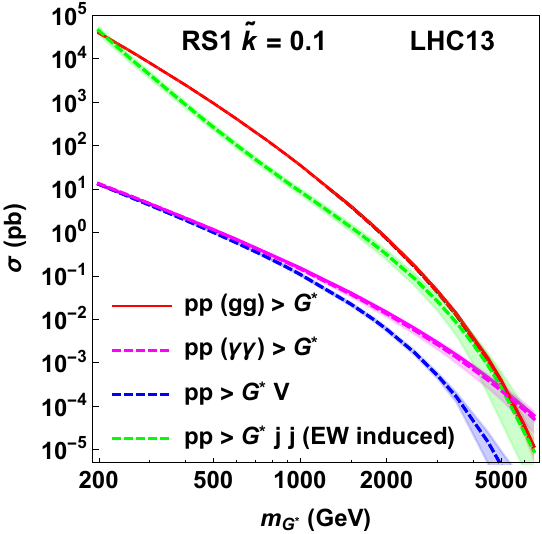}
\includegraphics[width=0.45\textwidth, angle =0 ]{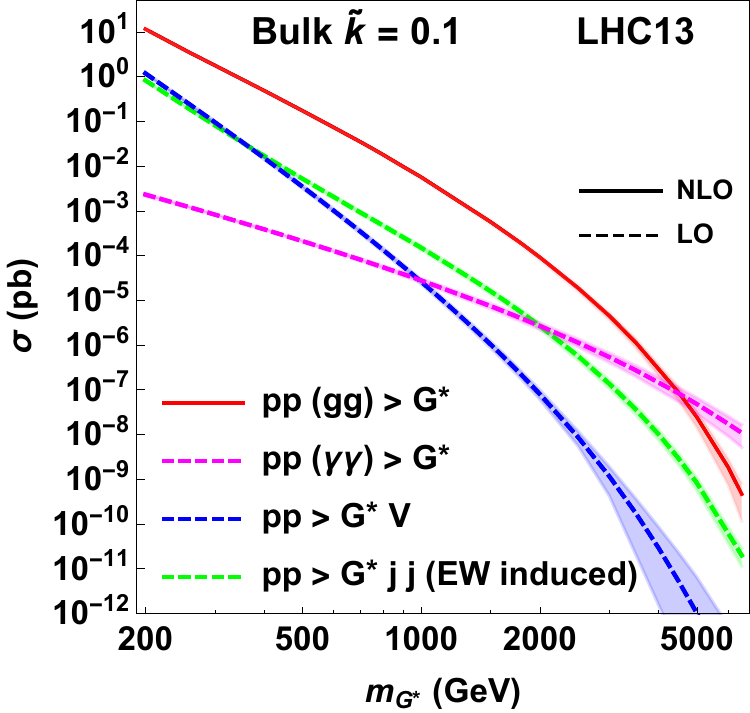}
\includegraphics[width=0.44\textwidth, angle =0 ]{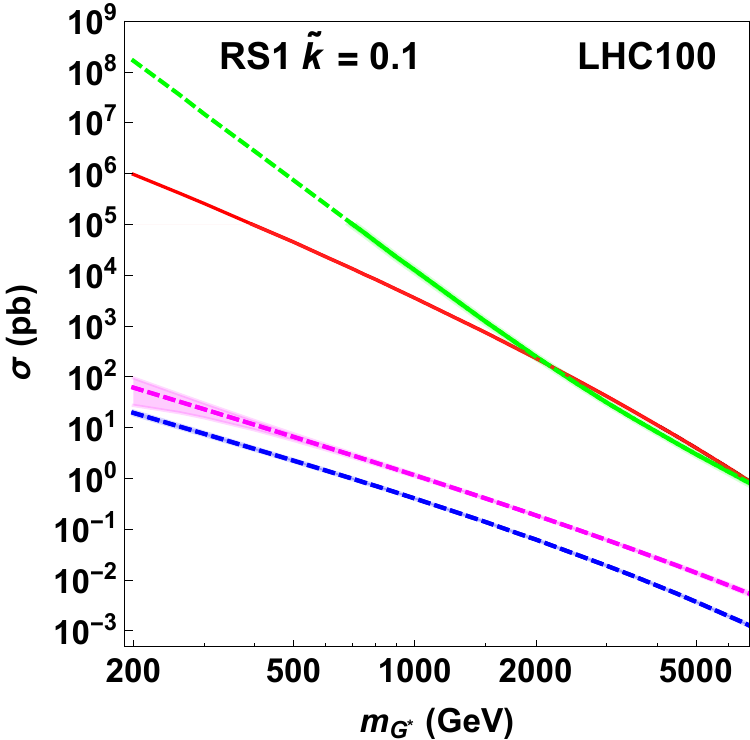}
\includegraphics[width=0.44\textwidth, angle =0 ]{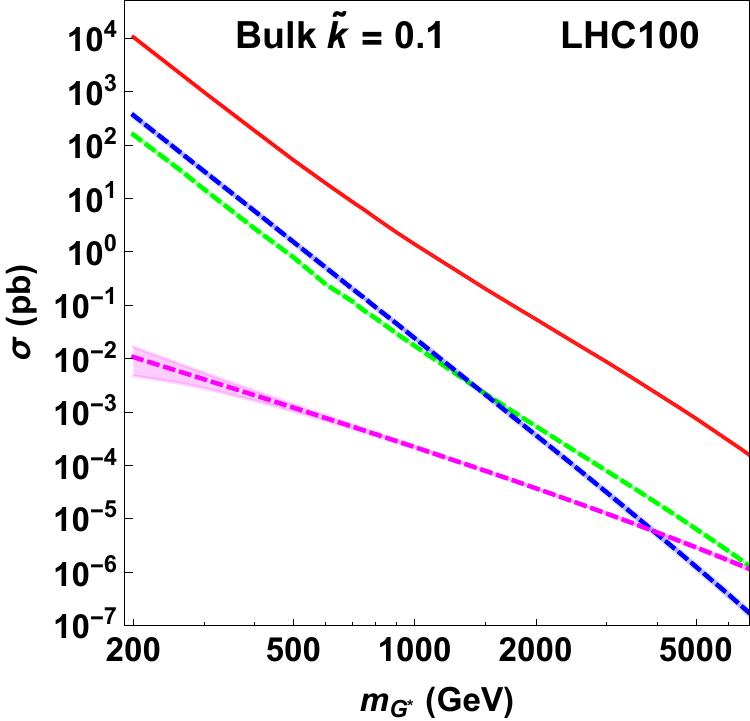}
\caption{\small{Production cross sections in pb for the KK-graviton with $k/M_{Pl} = 0.1$.  
The red curves correspond to the inclusive production, the green ones to the associated production with two jets, blue for associated production with a vector bosons and the magenta ones to photon-fusion production.
The NLO calculations are shown in continuous lines while the LO in dashed. When the NLO QCD results is available the uncertainty bands correspond to that calculation. 
{\bf Top:} 13 TeV. {\bf Bottom:} 100 TeV. 
{\bf Right:} Bulk scenario. {\bf Left:} RS1 scenario. 
}\label{figure:Gprod13}}
\end{center}\end{figure}

The production cross section of the $G^*$ in 13 TeV proton-proton collisions on RS1 and bulk scenario are shown in figure \ref{figure:Gprod13}. 
The GF is the dominant production mode for a $G^*$ resonance. The photo-production starts to be comparable with the QCD induced production only when $m_{G^*}$ is approximately a half of the total energy of the collision. 
The Gjj production have at least one order of magnitude smaller rate in the bulk scenario, and the GV production negligible. 
In the RS1 scenario Gjj starts to be comparable with the QCD induced production when $m_{G^*} >$ 2 TeV. 
The $\delta_{PDF}$ explodes to $m_{G^*} >$ 3 TeV. 

\begin{figure}[h]\begin{center}
\includegraphics[width=0.45\textwidth, angle =0 ]{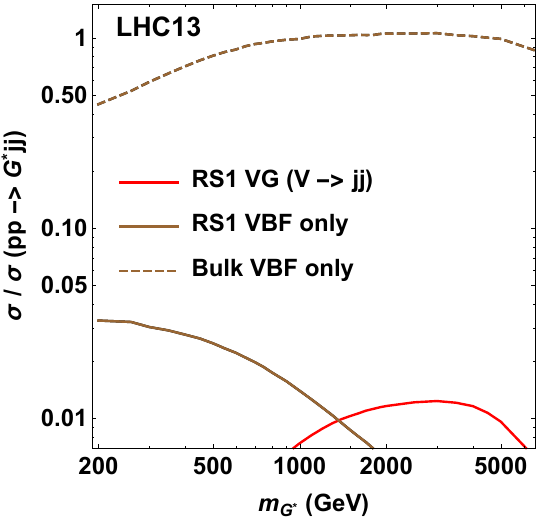}
\caption{\small{ 
Composition of the Gjj production cross section in sub-channels. 
The continuous curves stand for the RS1 scenario, where brown stand for the VBF-only component and red for the 
associated production with an hadronic vector boson. 
The dashed line stand for the VBF-only component  in the bulk scenario. 
}\label{figure:RS1prod}}
\end{center}\end{figure}

In figure \ref{figure:RS1prod} we show the composition of this signal by the ratio between the cross section for the different sub-processes simulated separately and the total $Gjj$ signal. 
We note that the pure VBF process (a) is dominant production mode to $m_{Gr} < \sim$400 GeV. When $m_{Gr} > \sim$400 GeV the qqV$G^*$ contact interaction (b and c) starts to dominate the ate, followed by the GV process (d+e+f). It is expected that the processes (b) and (c) dominates the total cross section, since it involves less propagators while the coupling strengths for the  qqV$G^*$ and VV$G^*$ interactions have the same order of magnitude and no momentun dependency. For the same reason the process (f) dominated the VG production at high energies.  
The study of spin-2 nature in VBF topology in the context of the Higgs boson characterization, with mass $\sim <$O(600) GeV, was made for example by references~\cite{Hagiwara:2009wt,Englert:2012xt}. It is not the scope of this paper to study the topology of the additional jets for the case of KK-gravitons with TeV scale masses.  

\subsubsection{Decay}

\begin{figure*}[h]\begin{center}
\includegraphics[width=0.45\textwidth]{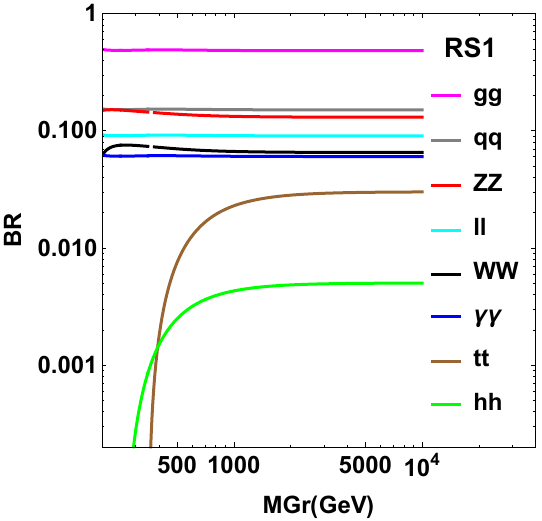}
\includegraphics[width=0.45\textwidth]{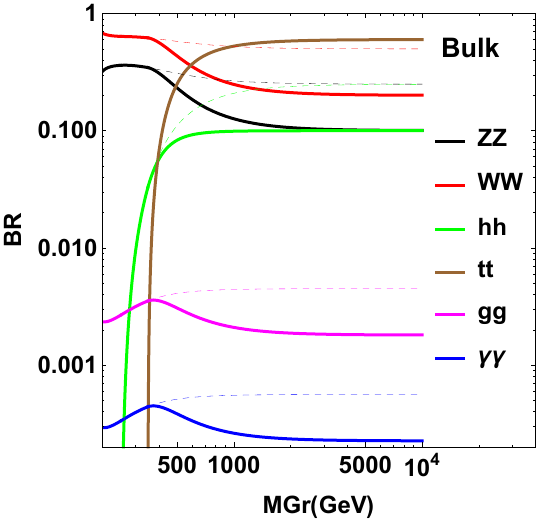}
\caption{ \small{
Branching fractions of $G^*$. 
{\bf Left:} RS1 scenario.  
The symbol $q$ stands for the sum of light quarks ($u,d,s,c,b$), while $l$ represents the sum of the three flavors of leptons ($e,\mu,\tau$) or neutrinos.
{\bf Right:} Bulk scenario comparing two hypothesis of fermion embedding. 
The branching ratios are independent of $\title{k}$ parameter. 
}}
\label{figure:ct}\end{center}\end{figure*}

On the left part of figure \ref{figure:ct} we show the $G^*$ branching fractions to SM particles. 
The left part of this figure shows the $G^*$ branching ratios on the RS1 scenario and the right part shows the bulk scenario in two cases: the thick curves considers a fully elementary top quark~\cite{Tuomas} while thin dashed ones we ignore $G^*$ coupling with top quark. 
In the RS1 scenario the highest branching fraction  is to di-jet final states. 
In the bulk case by other side (see the right part of figure \ref{figure:ct}) the highest branching fractions are to pairs of massive particles: weak bosons and top quarks. It is not surprisingly that $G^*$ branching ratios to pairs of massive bosons dominates once $d_g = 0$. 

\begin{figure*}[h]\begin{center}
\includegraphics[width=0.42\textwidth]{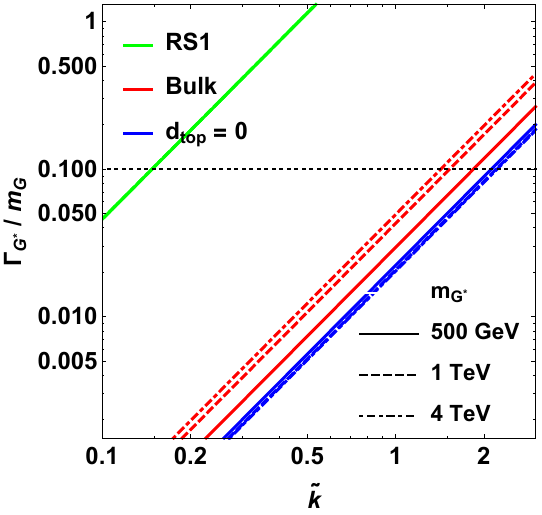}
\caption{\footnotesize{ Total width for a $G^*$ with $k/M_{Pl} = 0.1$. The green curve represents RS1 scenario and the red curves represents bulk scenario. We show two different hypothesis of treating $G^*$ couplings to bulk third family of quarks: the continuous curve considers a fully elementary top quark (\cite{Tuomas}) while dashed curve ignores $G^*$ coupling with top quark. 
} }
\label{figure:TW}\end{center}\end{figure*}

Figure~\ref{figure:TW} shows the $G^*$ total rate ($\Gamma_{G^*}$). 
As reflex of the bulk suppression of $G^*$ couplings to matter (see section~\ref{sec:couplings}) this is around two orders of magnitude larger in RS1 scenario in comparison with bulk scenario given the same geometric parameters. 
 
\begin{figure*}[h]\begin{center}
\includegraphics[width=0.45\textwidth]{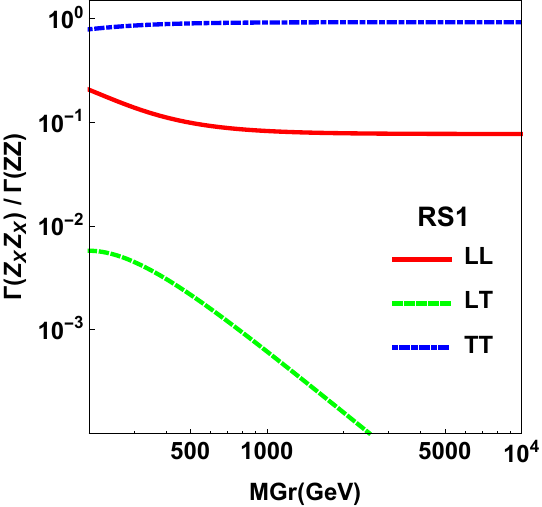}
\includegraphics[width=0.45\textwidth]{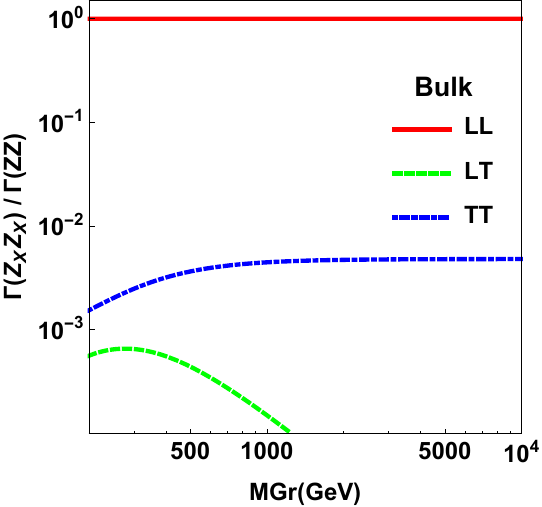}
\caption{\small Separation of polarizations on decay product of $h_{\mu\nu}^{(1)} \to ZZ$. The same behaviour holds to $h_{\mu\nu}^{(1)} \to WW$ decays. The continuous red line represents the portion of decay to pure longitudinal gauge boson modes (LL), the dot-dashed blue is the portion of decay to pure transverse modes (TT) and the dashed green the portion of decay to longitudinal-transverse mixed final states (LT).
}
\label{figure:pol}\end{center}\end{figure*}

For the case of a  resonance with $m_{G^*} >\sim 1$ TeV that decays to a boson pair, each one of the di-bosons products are boosted and consequently its sub-products collimated. The development of substructure techniques makes pairs of weak bosons to be golden channels for the search of $G^*$ in the bulk scenario. 
Boosted boson taggers can be  sensitive to the polarization of the (see for example \cite{JME-13-006}), therefore is phenomenologically 
interesting to understand the polarization of the vector bosons coming from the $G^*$ decay. 
Figure (\ref{figure:pol}) shows how its decay rate to weak bosons separates in  the different polarizations. 
As suggested by equation \ref{eq:GVV} in the RS1 scenario the $G^*$ decays to preferentially to transverse polarized modes while on bulk scenario it decays preferentially to longitudinally polarized modes, making those two benchmarks an excellent proxy to study the sensitivity of the vector boson taggers to their polarization. 

\clearpage
\subsection{Radion production and decay at LHC}

In many theories a scalar resonance is the most promising candidate to be the lightest particle on the particle spectra, as it is in WED~\cite{Csaki:2002gy}. 
As pointed out in the end of section~\ref{sec:radcoupl}, provided that the narrow width approximation is valid the phenomenology of a radion and a heavy Higgs is identical, with the only exception of the VBF and the VX production in bulk scenario. 

When applicable we will calculate the radion production cross sections making a simple rescaling of the state of the art calculations for a generic Higgs-like particle done in the context of the Higgs Cross section Working Group~\cite{MelladoGarcia:2150771,Denner:2047636}. 
 In this reference cross sections for the GF production mode are calculated at NNLO + NNLL order in QCD including soft-gluon resummation, as described in~\cite{deFlorian:2009hc}.  We ignore the NLO electroweak corrections provided. 
 The PDF used is PDF4LHC15@NNLO with 30 replicas, and the renormalization and factorization scales choices follows the description from the beginning of the chapter. 
The born level couplings between radion and massless bosons are:
\begin{subequations}
\label{eq:bornn}
\begin{equation}
\frac{c_{iir}}{\Lambda_R}(p_1\cdot p_2 g_{\mu\nu} - p_{1,\mu}p_{2,\nu})\,,
\end{equation}
where $i=g, \gamma$ and:
\label{eq:ggcoup}
\begin{equation}
c_{ggr} = \frac{\alpha_s}{4\pi}  \sum_{fermions} Q_f F_{1/2} (\tau_f)  -a_g \,
\end{equation}
\begin{equation}
c_{\gamma\gamma r} = \frac{\alpha_{QED}}{4\pi}\left(  \sum_{fermions} Q_f F_{1/2} (\tau_f)
+  F_{1} (\tau_w)\right) -a_{\gamma} \,,
\end{equation}
\end{subequations}
where $\tau_i = m_{R^*}/(4 m_i^2)$,  $F_{1/2}$ and $F_1$ are the one loop level functions to Higgs-like scalar coupling to massless bosons mediated by fermion and weak boson loops, as defined in~\cite{Spira:1995rr}, and $Q_f$ the charge of the fermion on the loop. 
The contribution from the top quark is the dominant in the loop function, however we also consider the contributions from the bottom and charm and tau running in the loop. 
The production cross sections of a  $X$ particle is matched with the one of  $R$ using the relation: 
\begin{equation}
\sigma_{R^*}^{GF} (m_{R^*}) = \frac{1}{\Lambda_R^2} \frac{ | c_{ggR^*} |^2}{|c_{SM}|^2}\times \sigma_{X}^{GF} (m_{X} = m_{R})\,, 
\label{eq:radionXS}
\end{equation}
where $c_{SM}= \frac{\alpha_s}{4\pi}  \sum Q_f F_{1/2} (\tau_f)$. 

The strength of the radion coupling with gluons is dominated by the point-like interaction induced by the $\beta_{QCD}$, 
that can be interpreted as take the infinite top approximation in the calculation on the GF of a Higgs-like boson.  
When  $m_X > 1$~TeV the infinite top approximation is valid in the calculation of the k-factor.
To $m_X$ closer to $2\times m_T$ the infinite top approximation is good up to 10\% in a NLO calculation~\cite{Kramer:1996iq,Catani:2003zt}. 
We also consider the PF production for the radion case, calculated at LO with .  
Again the point-like interaction proportional to $\beta_{QED}$ is dominant in the coupling strength.  
Conservatively we add 10\% uncertainty in both GF and PF production to include finite top mass effects in the soft gluon resumation. 

We calculate the radio VBF production including the effect of the bulk term (of equation~\ref{eq:rvvbulk}) in the  using the MG5 model implemented by~\cite{Maltoni:2013sma}. This implementation allows to calculate the NLO QCD VBF production of a generic scalar, taking into account the anomalous coupling between the scalar and the vector bosons induced by the bulk term. 
The model parameters used in the model~\cite{Maltoni:2013sma} the relevant parameters can be chosen as $c_{\alpha} =1$, $\kappa_{SM} = \frac{v}{\Lambda_R}$ and  $\Lambda = \Lambda_R$; 
In addition in the bulk scenario one should use $\kappa_{HWW} = \kappa_{HZZ} = \frac{1}{2 k \pi r_c}$ while $\kappa_{HWW} = \kappa_{HZZ} = 0$ in the RS1 scenario. 
In the radion case the VX  production with a is negligible in comparison with the VBF one, and therefore neglected~\cite{Denner:2047636}.

\begin{figure*}[h]\begin{center}
\includegraphics[width=0.45\textwidth, angle =0 ]{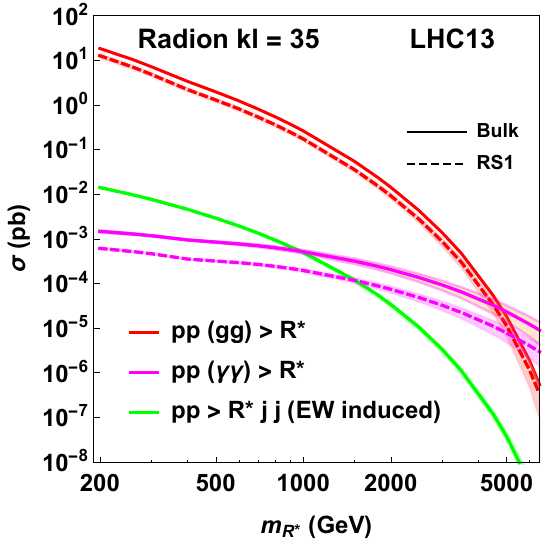}
\includegraphics[width=0.45\textwidth, angle =0 ]{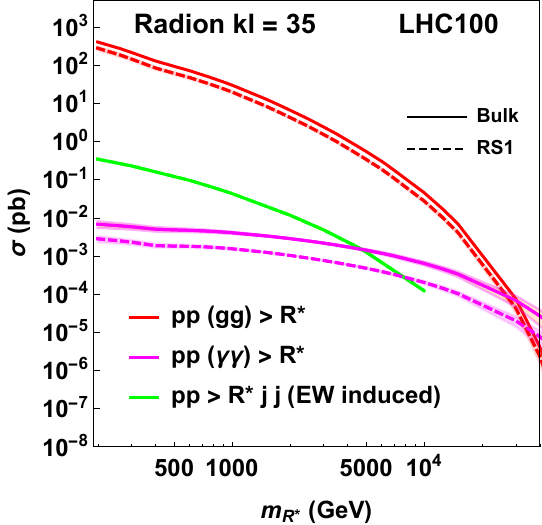}
\caption{\small{
 \small{Production cross sections in pb for the Radion, with $kl = 35$ and $\Lambda_R = 3$ TeV. The GF (red) and VBF (Green) production modes for both to bulk (continuous lines) and RS1 (dashed lines) scenarii.
 {\bf Right:} 13 TeV. {\bf Left:} 100 TeV. 
}
 }\label{figure:Radionprod13}}
\end{center}\end{figure*}

In figure~\ref{figure:Radionprod13} we show the production cross section for the Radion, with $kl = 35$ and $\Lambda_R = 3$ TeV and the effect of the bulk term in radion production. 
The continuous lines stand for the bulk scenario with $k r_c\pi = $35 while the dashed ones for the RS1 scenario. 
When $k r_c\pi = $35 the bulk terms increases the GF production rate by a factor $\sim$2 in the whole mass spectra. For the VBF production however the effect of the bulk term is negligible for the total cross section, what indicates that the bulk term also be negligible for differential distributions. It is not the scope of this note to study the kinematic properties of the VBF jets in the different hypotheses, for the case of RS1 radion (\~ CP even Higgs boson) this was done in~\cite{Maltoni:2013sma} and more recently for the 100 TeV case in~\cite{Goncalves:2017gzy}.
    
In the RS1 scenario the QCD NLO cross section for GF production is globally 10\%  lower than the approximate calculation at NNLO QCD accuracy done in the context of the HXSWG~\cite{MelladoGarcia:2150771,Denner:2047636}.

The radion total width ($\Gamma_R$) and branching ratios were extensively studied the literature, see for example~\cite{Csaki:2000zn,Csaki:2007ns,Barger:2011qn,Giudice:2000av,deSandes:2011zs}. We extend the analysis done in the literature to $m_R = 10$~TeV and analyze the phenomenological differences between the benchmarks considered.   
The radion branching ratios can be calculated with {\tt eHDecay}~\cite{Contino:2014aaa}, that is originally a tool to calculate decay rates and branching fractions of a Higgs-like particle using an effective lagrangian with dimension 6 operators. 
We use the  parametrization of the dimension 6 operators of the couplings in the SILH scheme~\cite{Contino:2013kra}. The calculations are done in NLO in QCD. 
The decay of the radion to a pair of 125~GeV Higgs bosons is added to the calculation at LO, following the equation bellow: 
\begin{equation}
\Gamma_{R'\to HH} = \frac{1}{32 \pi m_{R}^2}\sqrt{m_{R}^2-4 m_{H}^2}\frac{(m_{R}^2-2 m_{H}^2)^2}{\Lambda_R^2}
\end{equation}

\begin{figure*}[h]\begin{center}
\includegraphics[width=0.45\textwidth]{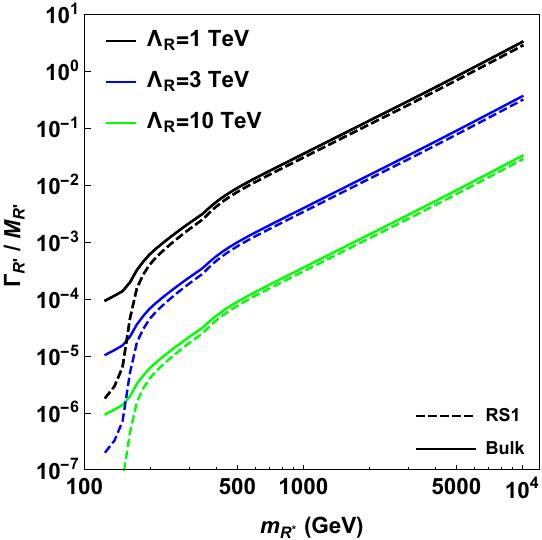}
\caption{{\small 
Radion total width and branching fractions. 
The continuous lines stand for the RS1 scenario while the dashed ones for the  bulk scenario with $k r_c\pi = $13.
}}
\label{figure:radionTW}\end{center}\end{figure*}

\begin{figure*}[h]\begin{center}
\includegraphics[width=0.45\textwidth]{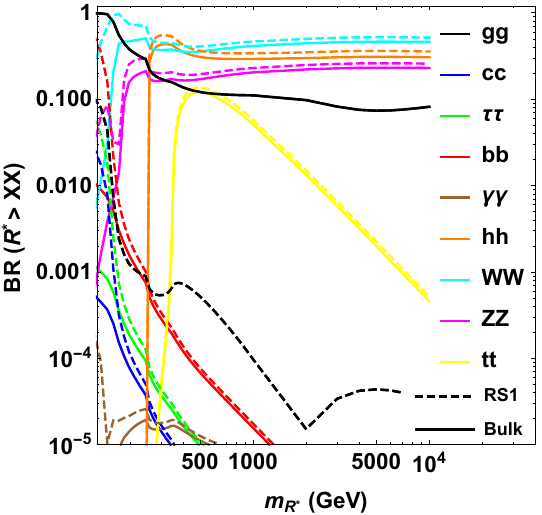}
\includegraphics[width=0.45\textwidth]{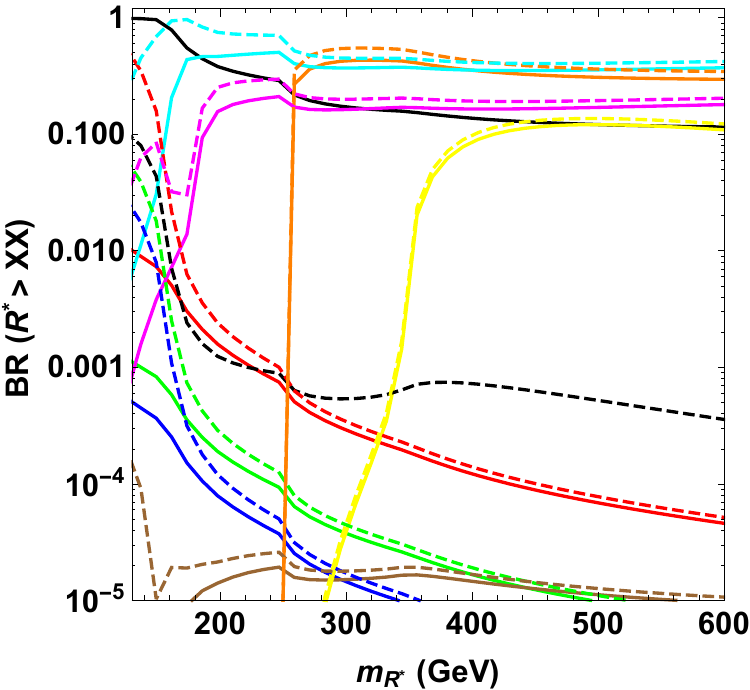}
\caption{{\small 
Radion branching fractions to bosons and fermions. The dashed lines stand for RS1 scenario, while the continuous ones for the bulk when $kl = $35. 
The figure on the left is a zoom of the figure on the right, following the same legend. 
}}
\label{figure:radionBR}\end{center}\end{figure*}

Figure~\ref{figure:radionTW} shows the ratio $\Gamma_R/m_R$, and figure~\ref{figure:radionBR} shows the radion branching ratios 
for the RS1 and bulk scenarios, in the last we chose $kl = $35 as benchmark.  
The total width ($\Gamma_R$) is inversely proportional to $\Lambda_R^2$ and the individual branching ratios are insensitive it. 
Generically its branching ratios to massive particles are dominant. 
The decay rate to bosons is proportional to $m_R^4$ while to fermions  is proportional to $m_R^2$ making the  branching ratio to a top quark pair to be suppressed to $m_R > $1~Tev. 
The sensitivity of the radion branching ratios to heavy particles the parameter $k r_c$ is negligible. 

\subsection{Summary and remarks}\label{sec:remarks}

\begin{figure*}[h]\begin{center}
\includegraphics[width=0.32\textwidth, angle =0 ]{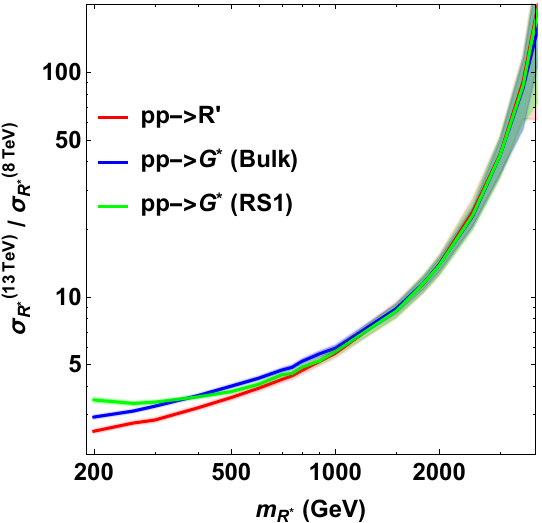}
\includegraphics[width=0.32\textwidth, angle =0 ]{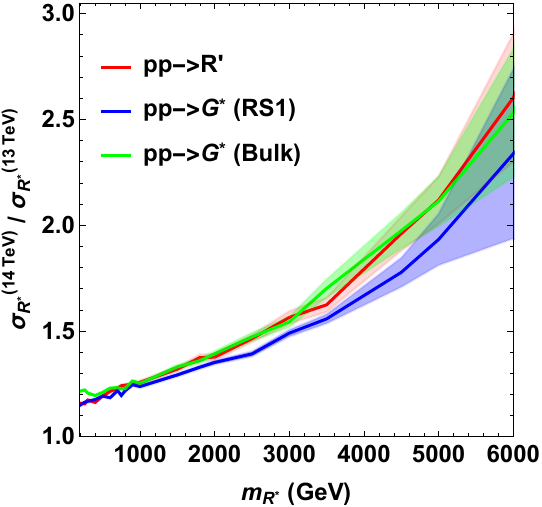}
\includegraphics[width=0.32\textwidth, angle =0 ]{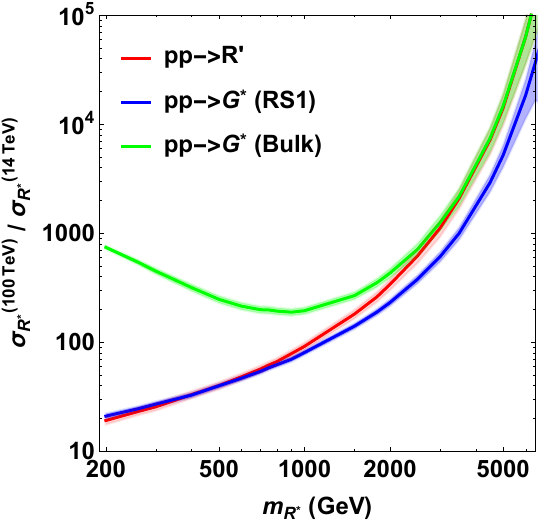}
\caption{\small{
Ratio of cross sections for QCD induced  production mode to different collider energies. 
}\label{figure:lumi}}
\end{center}\end{figure*}

In table \ref{fig:summaryXS} we summarize the production hypotheses, production modes and perturbation level that we had considered in this note. All the results derived in this note are made public in electronic format for direct use of by the experiments~\cite{WEDgit}. 
 In appendix \ref{app:tables} we write a practical guide on how to variate model parameters in the different benchmarks. 
 In figure~\ref{figure:lumi} we show the enhancement of the QCD induced production cross section when 
 we change the energy of LHC from 8~TeV to 13~TeV, from 13~TeV to 14~TeV and from 14~TeV to 100~TeV. 
  The gain in cross section when increasing the LHC energy from 14~TeV to 100~TeV is larger  for the radion and the KK-graviton in the bulk scenario, that are purely GF produced. 

\begin{table}
\centering
{\small
\begin{tabular}{ |cc|c|c|c|c|c| } 
\hline
& & QCD induced & PF & Xjj (EW induced) & XV & Decay \\\hline
radion & RS1 & NNLO + NNLL & LO & NLO & - & NLO \\
           & Bulk & (NLO) &  & & & \\\hline  
KK-graviton & RS1 & NLO & LO & \multicolumn{2}{c|}{ LO } & LO\\ 
                   & Bulk &    &    & \multicolumn{2}{c|}{ \mbox{}} &  \\\hline
\end{tabular}
}
\caption{\small{
Summary of the cross section results, highlighting at which QCD perturbative order the calculations are available. 
When the level where events can be simulated is different from the level where the cross section is calculated this is 
showed between parenthesis. The boxes are merged when the models and tools used are the same.  
}}
\label{fig:summaryXS}
\end{table} 

\section*{Conclusions}
\markboth{Conclusions}{}%
\addcontentsline{toc}{section}{Conclusions}

Extensions of the SM based on the existence of extra dimensions are predictive scenarios that motivates the search for high mass resonances with spin even (0 or 2) particles at LHC. The predictions of such class of theories for the 
effective couplings of resonances illustrate the phenomenology that is relevant to collider searches. 
We had reviewed the physics of a heavy spin-2 particle interpreting it as KK-graviton under Warped Extra Dimension scenario. Two hypothesis was exploited: RS1 (where only gravity is allowed to propagate on the extra dimensional bulk) and a bulk scenario where SM matter is also allowed to propagate on the extra dimensional direction, using the same gauge construction of SM and no additional model building hypotheses. 

In light of the above defined benchmarks we had calculated its cross sections, total width and branching ratios  using the most up-to-date techniques available and put side by side the results for RS1 and bulk scenarios and allow a fair comparison between the hypotheses. 
On the course of calculation we had cross checked all the results with or analytic calculations, and/or alternative Monte Carlo implementations and/or with other public results available in the community.

\section*{Acknowledgements}

This note would not be possible without the fruitful physics discussions with Tuomas Hapola, Kaustubh Agashe, Hooman Davoudiasl, Gilad Perez and Veronica Sanz. The note would also not be possibe without the idea from Alessio Bonato, Maurizio Pierini, Francesco Santanastasio and Thiago Tomei, that fully supported the initial cross checks between the different Monte Carlo implementations. 
I am very grateful to Eduardo Ponton, Maxime Gouzevich and Angelo Santos for the careful reading and the valuable comments on both physics and text. And also to Sasha Belyaev for the last comment on parameter space.
I am also grateful Miguel and Mateus, for forcing me do the night shifts taking care of them and writing this note. .C. is supported by MIURFIRB RBFR12H1MW grant. 

\begin{appendices}
\section{Decay rates formulae}

For convenience we rewrite here the analytic expressions to the $G^*$ decay rates derived in \cite{Han:1998sg}, 
adapted to the benchmark models we consider in this document. 
The decay rate of $G^*$ to a pair of physical Higgses is:
\begin{eqnarray}
\Gamma (h_{\mu\nu} \to HH) = \frac{m_G^3}{960\pi}\left( \frac{\tilde{k} x_1}{m_G}\right)^2\left(1 -\frac{4 m_H^2}{m_G^2} \right)^{5/2} 
\end{eqnarray}
The decay rate of the $G^*$ to a pair of SM non-massive vectors is given by equation:
\begin{subequations}
\begin{equation}
\Gamma (h_{\mu\nu} \to gg, \gamma\gamma) = 2 d_{g}^2 \times N_{g,\gamma}\times \frac{m_G}{160\pi} \left( \tilde{k} x_1 \right)^2\,,
\end{equation}
\begin{equation}
N_{g} = 8 \, , N_{\gamma} = 1\,. 
\end{equation}
\label{eq:iden} 
\end{subequations}
The decay rates of $G^*$ to a pair of W bosons is:
\begin{subequations}
\begin{eqnarray}
\Gamma (h_{\mu\nu} \to ZZ) = \frac{m_G^3}{960\pi} \left( \frac{\tilde{k} x_1}{m_G}\right)^2 ( A_Z + d_g B_Z + d_g^2 C_Z)  \sqrt{1-4\frac{m_i^2}{m_G^2}},
\end{eqnarray}
\begin{equation}
\Gamma (h_{\mu\nu} \to WW) = 2\times \Gamma (h_{\mu\nu} \to ZZ) 
\end{equation}
\end{subequations}
where for each $i=W,Z$:
\begin{subequations}
\begin{eqnarray}
A_i = \left( 1 + 12\frac{ m_i^2}{m_G^2} + 56\frac{ m_i^2}{m_G^2}\right)  
\end{eqnarray}
\begin{equation}
B =  80 \left( 1 -  \frac{m_i^2}{m_G^2} \right) \frac{ m_i^2}{m_G^2} 
\end{equation}
\begin{equation}
C =  12 \left( 1 -  3\frac{m_i^2}{m_G^2} + 6\frac{m_i^4}{m_G^4} \right)\,.  
\end{equation}
\label{eq:Vpol}
\end{subequations}
The decay rates of $G^*$ to a pair of W bosons is:

The general expression for the decay into a pair of light fermions (where we consider $c_{f_L} = - c_{f_R} \equiv c_f$)  with mass $m_f$ is:
\begin{eqnarray}
\Gamma (h_{\mu\nu} \to f\bar{f}) = \frac{m_G^3}{80 \pi} \left( \frac{\tilde{k} x_1}{m_G}\right)^2 
\left(  1-\frac{8}{3}\frac{m_f^2}{m_G^2} \right) \left( 1-4\frac{m_f^2}{m_G^2} \right)^{\frac{3}{2}}\,.
\end{eqnarray}
For the case of the top quark we consider the elementary top hypothesis. The decay rate of $G^*$ to a pair of top quarks with mass $m_T$ is:

\begin{eqnarray}
\Gamma (h_{\mu\nu} \to t\bar{t}) = \frac{m_G^3}{80 \pi} \left( \frac{\tilde{k} x_1}{m_G}\right)^2 
\left(\frac{3}{4} - \frac{7}{2} \frac{m_T^2}{m_G^2} +2\frac{m_T^4}{m_G^4} \right) \sqrt{1-4\frac{m_T^2}{m_G^2}}
\end{eqnarray}

\section{Modifying model parameters}
\label{app:tables}

To the KK-graviton in the RS1 case, and as well for bulk case in the EW induced production mode and to and to all Radion production modes the cross sections scale respectively only with $\tilde{k}$ or $\Lambda_R$ in the following manner:
\begin{equation}
\sigma (m_{G^*},\tilde{k}) =  \left(\frac{\tilde{k}}{0.1}\right)^2 \sigma (m_{G^*}, \tilde{k} = 0.1)\,.
\label{eq:cxscaling}
\end{equation}

\begin{equation}
\sigma (m_{R},\tilde{k}) =  \left(\frac{3\mbox{ TeV}}{\Lambda_R}\right)^2 \sigma (m_{R}, \Lambda_R = 3\mbox{ TeV})\,.
\label{eq:cxscaling}
\end{equation}

The branching ratios of $G^*$ are not dependent of $\tilde{k}$. Its total decay rate however also scales with $\tilde{k}$ in the same form of equation \ref{eq:cxscaling}. 
To the gluon fusion and photon fusion  production modes of a $G^*$ in bulk scenario we need to take into account the 
mild dependence of $\tilde{k}$ in the coupling volume suppression $d_g$:
\begin{equation}
\sigma_{bulk}^{(gg/\gamma\gamma)} [m_{G^*},\tilde{k}] = \left(  \frac{d_g (m_{G^*}, \tilde{k})}{d_g(m_{G^*}, \tilde{k} = 0.1)} \times \frac{\tilde{k}}{0.1}\right)^2 \sigma_{bulk}^{ (gg/\gamma\gamma)} [m_{G^*}, \tilde{k} = 0.1]\,.
\end{equation} 
The photo-production in the bulk scenario and RS1 scenario are related by a simple rescaling:
\begin{equation}
\sigma_{bulk}^{(\gamma\gamma)} [m_{G^*},\tilde{k}] = \left(  d_g (m_{G^*}, \tilde{k})\times \tilde{k} \right)^2 \sigma_{RS1}^{ (\gamma\gamma)} [m_{G^*}, \tilde{k} = 0.1]\,,
\end{equation}

\end{appendices}
\printbibliography

\end{document}